\documentclass[%
 aip,
 amsmath,amssymb,
 reprint,%
]{revtex4-1}

\usepackage{graphicx}
\usepackage{dcolumn}
\usepackage{bm}
\usepackage{diagbox} 
\usepackage{booktabs}
\usepackage{physics}
\usepackage[colorlinks=true, linkcolor=blue, citecolor=blue, urlcolor=blue]{hyperref}
\usepackage{subcaption}
\usepackage{ragged2e} %
\usepackage[format=plain,justification=justified,singlelinecheck=false]{caption}

\makeatletter
\def\@email#1#2{%
 \endgroup
 \patchcmd{\titleblock@produce}
  {\frontmatter@RRAPformat}
  {\frontmatter@RRAPformat{\produce@RRAP{*#1\href{mailto:#2}{#2}}}\frontmatter@RRAPformat}
  {}{}
}%
\makeatother
\begin{document}


\title{
Sensor free, self regulating thermal switching \\
via anomalous Ettingshausen effect and spin reorientation in DyCo$_{5}$
}
\author{Shibo Wang}
\author{Hiroki Tsuchiura}
 \thanks{Author to whom correspondence should be addressed: tsuchi@tohoku.ac.jp}
 \affiliation{Department of Applied Physics, Tohoku University, Aoba, Sendai 980-8579, Japan\linebreak}
\author{Nobuaki Terakado}
\affiliation{Department of Applied Physics, Tohoku University, Aoba, Sendai 980-8579, Japan\linebreak}
\affiliation{Department of Material Chemistry, Kyoto University, 
  Katsura, Nishikyo-ku, Kyoto 615-8520, Japan\linebreak}


\begin{abstract}
We propose a sensor free, self regulating thermal switch that combines the anomalous Ettingshausen effect (AEE) with a temperature driven spin reorientation transition (SRT) in the rare earth cobalt compound DyCo$_5$.
Using density functional theory and the Kubo linear-response formalism, we compute the anomalous Hall conductivity $\sigma_{xy}(\varepsilon)$ and the finite temperature anomalous Nernst conductivity $\alpha_{xy}(T)$ for two magnetization directions, $\bm{M}\parallel c$ and $\bm{M}\perp c$.
While the intrinsic $\sigma_{xy}$ at the Fermi level remains sizable for both orientations, $\alpha_{xy}$ exhibits about two orders of magnitude contrast in the SRT temperature window.
This contrast is consistent with the low temperature Mott relation through the energy slope $\partial_\varepsilon \sigma_{xy}(\varepsilon)\rvert_{E_{\mathrm F}}$ and is traced to strongly peaked Berry curvature hot spots generated by spin orbit coupling induced avoided crossings of Co $3d$ bands.
Combining $\alpha_{xy}$ with longitudinal transport coefficients, we estimate device level metrics, namely the anomalous Nernst thermopower $S_{\mathrm{ANE}}$ and the Ettingshausen coefficient $\Pi_{\mathrm{AEE}}=T S_{\mathrm{ANE}}$, and demonstrate robust orientation controlled switching under a fixed in plane bias current.
These results establish a materials based route to compact thermal control without external sensors or feedback electronics and provide a concrete example that the proposed principle can be realized in an existing ferromagnet.
%
\end{abstract}

\maketitle

Modern microelectronic and energy systems increasingly require thermal control that is compact, localized, and as rapid as practicable~\cite{Pop2010,Moore2010}.
These requirements point to electrically driven mechanisms that are readily integrable on chip.
Among such mechanisms, transverse thermoelectric effects (TTEs)~\cite{Uchida2021}, notably the anomalous Nernst effect (ANE)~\cite{Nernst1887} and anomalous Ettingshausen effect (AEE)~\cite{Etting1886}, are attractive because they can operate without externally applied magnetic fields (e.g., field coils), once a ferromagnetic state provides spontaneous magnetization, and are governed by the Berry curvature of ferromagnets~\cite{Nagaosa2010}.
Their magnitudes, and even signs, are highly sensitive to the magnetization orientation, thereby providing a direct, materials-intrinsic means for functional control.
However, most switching concepts still rely on external sensors or feedback electronics to determine when to reverse the heat flow.
Unlike approaches where the magnetization is actively reoriented by an externally applied magnetic field, strain, or current-induced torques, the present concept uses an intrinsic temperature-driven spin reorientation transition (SRT) as the internal ``trigger.'' This makes the switching threshold materials-defined and enables sensor-free negative feedback: once the device temperature enters the SRT interval, the equilibrium $\bm{M}$ rotates and the AEE heat flux is redirected without external sensing or control circuitry.

In this context, a large AEE has recently been reported in SmCo$_5$-type permanent magnets~\cite{Miura2019}.
These results motivate exploring related $R$Co$_5$ compounds with additional internal control parameters, including DyCo$_5$ which exhibits a temperature-driven spin-reorientation transition \cite{Tsushima1983,Alena2025}.

\begin{figure}[h]
  \begin{center}
    \includegraphics[width=1\linewidth, height=7cm]{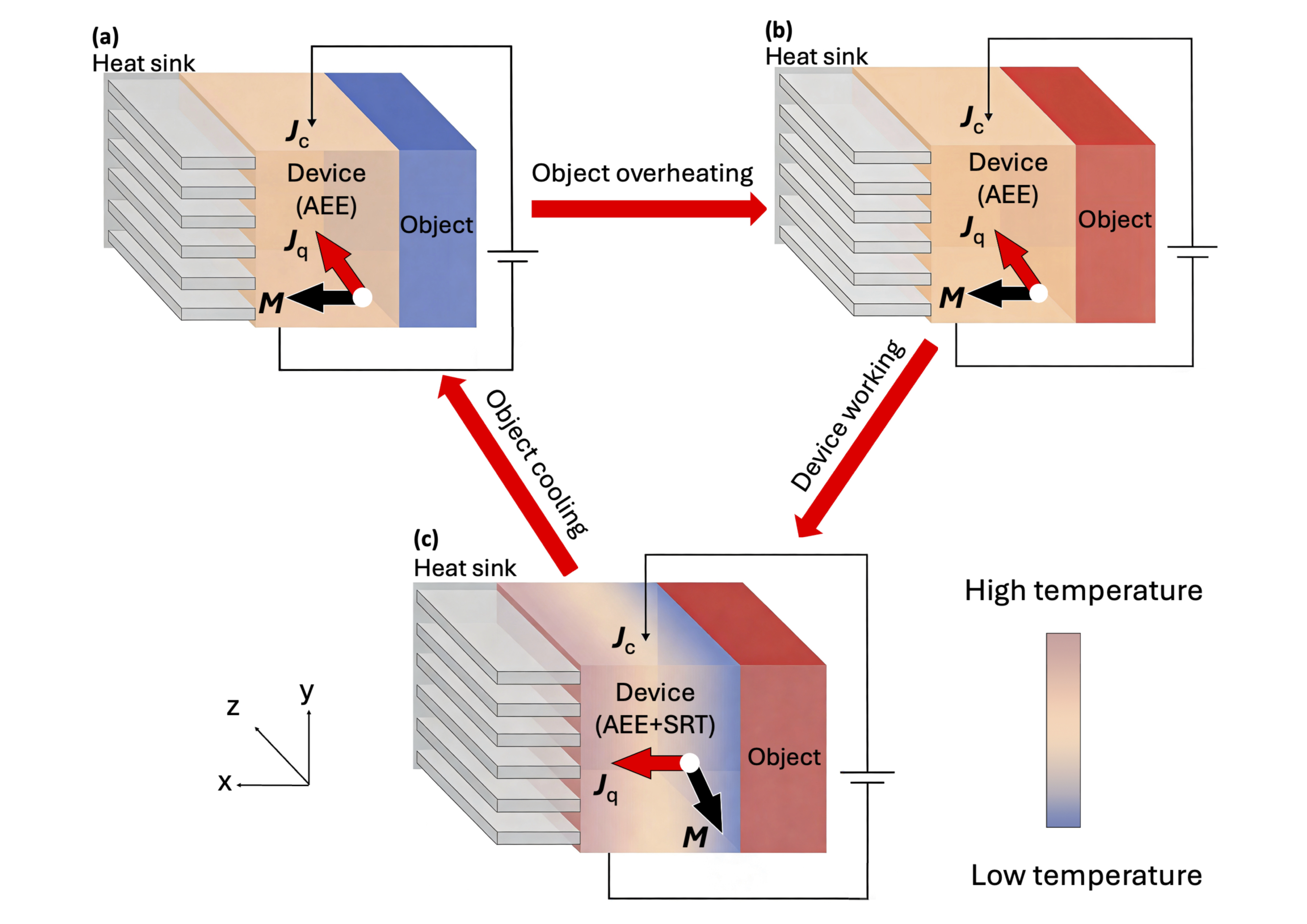}
    \caption{\justifying
    Schematic of a self regulating thermal control device based on the anomalous Ettingshausen effect (AEE) and a spin reorientation transition (SRT). Here, $T$ denotes temperature, $\bm{M}$ magnetization, $\bm{J}_c$ charge current density, and $\bm{J}_q$ heat current density. Under a constant in-plane $\bm{J}_c$, AEE generates a transverse $\bm{J}_q$ whose direction is set by $\bm{M}$. If the object is heated by $\delta T$, the device warms accordingly and, upon entering the SRT interval, $\bm{M}$ reorients ($\perp z \leftrightarrow \parallel z$), which redirects or reverses $\bm{J}_q$. This provides a negative-feedback response: heat is drawn from the object in the overheated state and the system relaxes back toward the baseline temperature without external sensing.
    } \label{fig:schema}
  \end{center}
\end{figure}

Herein, we propose a sensor-free, self-regulating thermal switch that leverages the AEE in DyCo$_5$, a member of the $R$Co$_5$ family that is known to exhibit an SRT~\cite{Ohkoshi1977,Tsushima1983,Tereshina2007}.
In this material, the magnetic easy axis rotates with temperature due to competing crystal-field and exchange anisotropies~\cite{Belov1976,Davila2024}; it lies in-plane for $T<T_{\mathrm{SR1}}\approx 325\,\mathrm{K}$ and switches to the $c$ axis at higher temperatures (with $T>T_{\mathrm{SR2}}\approx 367\,\mathrm{K}$)~\cite{Ohkoshi1977,Tsushima1983}.
As $T$ passes through the SRT interval, $T_{\mathrm{SR1}}\lesssim T \lesssim T_{\mathrm{SR2}}$, the magnetization $\bm{M}$ reorients from in-plane toward the $c$ axis.
Under a fixed in-plane bias current $\bm{J}_{c}$, this reorientation toggles the transverse anomalous Nernst conductivity $\alpha_{xy}$ and thus the AEE-induced heat flux $\bm{J}_{q}\propto\alpha_{xy}$, enabling temperature-thresholded directional switching or modulation of heat flow without any external sensing.
A schematic of the device concept is shown in Fig.\ref{fig:schema}.

To establish the mechanism and quantify performance, we combined density-functional calculations and the Kubo linear-response formalism to evaluate the energy-resolved anomalous Hall conductivity $\sigma_{xy}(\varepsilon)$ and the finite-temperature $\alpha_{xy}(T)$. We found that although the intrinsic $\sigma_{xy}$ at $E_{\mathrm F}$ remains sizable for both magnetization orientations of $\bm{M}$, $\alpha_{xy}$ changes by several orders of magnitude across the SRT. This contrast is consistent with the low-$T$ Mott relation through the energy slope $\bigl(\partial_{\varepsilon}\sigma_{xy}\bigr)\rvert_{E_{\mathrm F}}$~\cite{Mott1936,Mott1969}, reflecting orientation-driven redistribution of Berry curvature near $E_{\mathrm F}$. Device-relevant figures of merit, namely the ANE thermopower $S_{\mathrm{ANE}}$ and the Ettingshausen coefficient $\Pi_{\mathrm{AEE}}=T\,S_{\mathrm{ANE}}$~\cite{Seki2018}, exhibit robust, orientation-controlled switching at a fixed temperature (e.g., $300\,\mathrm{K}$), highlighting a materials-integrated route to compact, current-driven, sensor-free thermal control.



\noindent\textbf{Theoretical methods.}
We performed density functional theory (DFT) calculations using the full-potential linearized-augmented plane-wave (FP-LAPW) method~\cite{blaha2001wien2k}.
Spin--orbit coupling (SOC) is treated self-consistently within the second-variational FP-LAPW scheme.
Unless otherwise noted, the rare-earth $4f$ states are treated in the generalized gradient approximation (GGA)~\cite{Perdew1996} with a Hubbard correction $U$~\cite{Liechtenstein1995}, and a moderate on-site interaction on Co $3d$ is used only to test the robustness of trends.
We utilized the Wannier90 program~\cite{Pizzi2020} to construct maximally localized Wannier functions (MLWFs), which were then employed for Wannier interpolation to evaluate the Berry curvature.
MLWFs~\cite{Marzari2012} are constructed primarily from Co $3d$ and $s$ states (with rare-earth $f$) to reproduce the DFT bands.
Longitudinal thermoelectric properties were computed within semiclassical Boltzmann transport theory as implemented in BoltzTraP2~\cite{Madsen2018}.
In the present work, Wannier90 (MLWF interpolation) is used to construct an MLWF-based tight-binding model and analyze the Berry curvature, while $\sigma_{xy}$ and $\alpha_{xy}$ are obtained from Brillouin-zone (BZ) integrals of the Kubo expressions in WIEN2k. BoltzTraP2 is used only for $\sigma_{yy}/\tau$ and $S_{yy}$ within the constant relaxation time approximation required for Eq.~\eqref{eq:sane_smallangle}.
Two magnetization configurations were considered throughout: $\bm{M}\!\parallel\!c$ and $\bm{M}\!\perp\!c$ (in-plane).
BZ quantities were evaluated on a dense $68\times68\times73$ $k$-mesh; all energy-resolved and finite-$T$ integrals were examined for $k$-mesh convergence.

The intrinsic anomalous Hall conductivity (AHC) and anomalous Nernst conductivity (ANC), denoted by $\alpha_{xy}$, are computed from the $k$-space Berry curvature within the Kubo linear-response formalism:
\begin{equation}
  \sigma_{xy} \;=\; -\frac{e^2}{\hbar}
  \sum_{n}\!\int_{\mathrm{BZ}}\!\frac{d^3\bm{k}}{(2\pi)^3}\;
  f_{n\bm{k}}\,\Omega_{n,z}(\bm{k}),
  \label{eq:ahc_def}
\end{equation}
\begin{equation}
  \alpha_{xy} \;=\; \frac{e}{\hbar}
  \sum_{n}\!\int_{\mathrm{BZ}}\!\frac{d^3\bm{k}}{(2\pi)^3}\;
  \Omega_{n,z}(\bm{k})\, s_{n\bm{k}},
  \label{eq:anc_def}
\end{equation}
where $f_{n\bm{k}}=f\!\bigl(\varepsilon_{n\bm{k}}\bigr)$ is the Fermi--Dirac factor and
$s_{n\bm{k}} = -k_{\mathrm B}\!\left[f_{n\bm{k}}\ln f_{n\bm{k}} + (1-f_{n\bm{k}})\ln(1-f_{n\bm{k}})\right]$ is
the entropy density per state~\cite{Xiao2006,Xiao2010}.
The Berry curvature follows from
\begin{equation}
  \bm{\Omega}_n(\bm{k}) \;=\; \nabla_{\bm{k}}\times \bm{A}_n(\bm{k}),
  \qquad
  \bm{A}_n(\bm{k}) \;=\; i\,
  \big\langle u_{n\bm{k}}\big|\nabla_{\bm{k}}\big|u_{n\bm{k}}\big\rangle ,
  \label{eq:berry_defs}
\end{equation}
where $\lvert u_{n\bm{k}}\rangle$ are cell-periodic Bloch states.
Equations~\eqref{eq:ahc_def}--\eqref{eq:berry_defs} fix our sign and derivative conventions.

For interpretability we also employ the energy resolved AHC representation,
\begin{equation}
  \alpha_{xy} \;=\; -\frac{1}{eT}\int d\varepsilon\;
  \Bigl(-\frac{\partial f}{\partial \varepsilon}\Bigr)\,
  (\varepsilon-\mu)\,\sigma_{xy}(\varepsilon),
  \label{eq:alpha_energy}
\end{equation}
which reduces in the low temperature limit ($k_B T\ll E_{\mathrm F}$) using the Sommerfeld expansion to the Mott relation:
\begin{equation}
  \alpha_{xy} \;\simeq\; -\frac{\pi^2 k_B^2 T}{3e}\,
  \left.\frac{\partial \sigma_{xy}(\varepsilon)}{\partial \varepsilon}\right|_{\varepsilon=E_{\mathrm F}}.
  \label{eq:mott}
\end{equation}
Equation~\eqref{eq:mott} is used as an \emph{intuition aid}; all quantitative results are obtained from Eqs.~\eqref{eq:anc_def} and \eqref{eq:alpha_energy} at finite $T$.
We next connect $\alpha_{xy}$ to device relevant thermoelectric metrics.

Device relevant metrics, the transverse thermopower due to ANE and the Ettingshausen coefficient, are related by the Kelvin relation
$\Pi_{\mathrm{AEE}}=T\,S_{\mathrm{ANE}}$.
Under open circuit conditions in the transverse direction and for a small Hall angle, we use~\cite{Ikhlas2017}
\begin{equation}
  S_{\mathrm{ANE}}
  \;\approx\; \frac{\alpha_{xy}}{\sigma_{yy}} \;-\; S_{yy}\,\frac{\sigma_{xy}}{\sigma_{yy}},
  \label{eq:sane_smallangle}
\end{equation}
where $\sigma_{yy}$ and $S_{yy}$ are the longitudinal conductivity and Seebeck coefficient along the current direction.
While AHC and $\alpha_{xy}$ are determined by the Berry curvature in our formulation, the longitudinal conductivity depends on scattering.
We therefore compute $\sigma_{yy}/\tau$ and $S_{yy}$ using BoltzTraP2 within the constant relaxation time approximation.
In this approximation $S_{yy}$ is insensitive to the absolute value of $\tau$, whereas $\sigma_{yy}$ requires $\tau$.
We estimate $\tau$ by comparing calculated $\sigma_{yy}/\tau$ with reported experimental conductivities of $R$Co$_5$ compounds, and adopt $\tau=11.3$~fs and $\tau=12.7$~fs as representative values.
Representative values are summarized in Table~\ref{tab:tau}.
We report $S_{\mathrm{ANE}}$ and $\Pi_{\mathrm{AEE}}$ for both values to indicate the uncertainty associated with $\tau$.
We note that $\tau$ can be temperature dependent and may be further reduced near the SRT due to enhanced magnetic fluctuations and spin-disorder scattering.
Such effects primarily rescale the longitudinal conductivity entering Eq.~\eqref{eq:sane_smallangle}, whereas the key on/off contrast discussed here originates from the intrinsic $\alpha_{xy}$ computed from the Berry-curvature formulation.

\begin{table}
\caption{\label{tab:tau}Relaxation times used in the constant relaxation time approximation. The conductivity $\sigma$ values are taken from Refs.~\cite{Miura2019,Seifert2021}, and the present table was compiled by the authors (it is not reproduced or adapted from any published table).}
\begin{ruledtabular}
\begin{tabular}{ccc}
System & $\tau$ (fs) & Basis \\ \hline
SmCo$_5$ (bulk) & 11.3 & $\sigma$ taken from Ref.~\cite{Miura2019} and our $\sigma/\tau$ \\
DyCo$_5$ (thick film) & 12.7 & $\sigma$ taken from Ref.~\cite{Seifert2021} and our $\sigma/\tau$ \\
\end{tabular}
\end{ruledtabular}
\end{table}

We report AHC in S/m (1~S/cm $=$ 100~S/m), the anomalous Nernst conductivity $\alpha_{xy}$ in A\,m$^{-1}$\,K$^{-1}$, $S_{\mathrm{ANE}}$ in $\mu$V/K, and $\Pi_{\mathrm{AEE}}$ in $\mu$V.
Energies are referenced to $E_{\mathrm F}$ (shown as $E_{\mathrm F}\!=\!0$ in energy resolved plots).
Colors and line styles are maintained identical across panels to distinguish $\bm{M}\!\parallel\!c$ and $\bm{M}\!\perp\!c$.

\begin{figure}[htbp]
  \begin{center}
    \includegraphics[width=65mm]{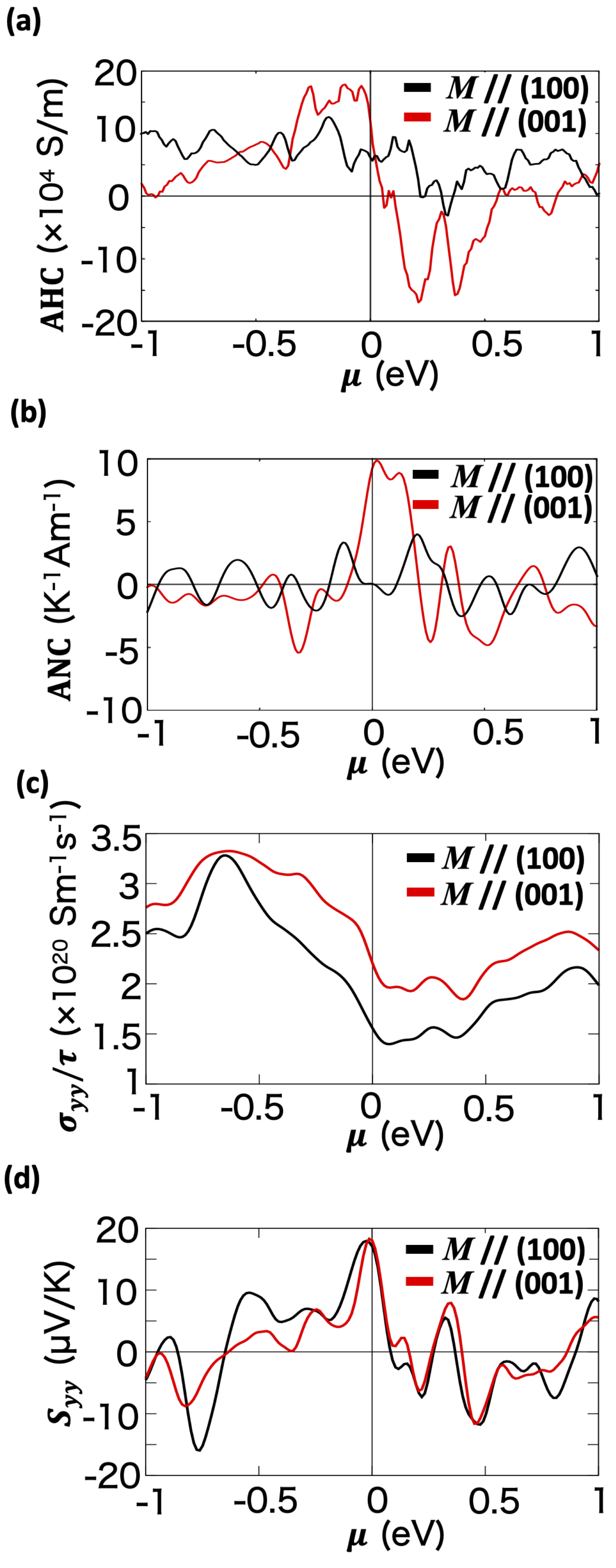}
    \caption{
    Transport coefficients of DyCo$_5$ as functions of the chemical potential $\mu$ for two magnetization directions, $\bm{M}\parallel(100)$ and $\bm{M}\parallel(001)$.
    The chemical potential is varied within a rigid band approximation to indicate the carrier doping trend, and $\mu=0$ corresponds to the Fermi level ($E_F$).
    (a) Anomalous Hall conductivity (AHC) $\sigma_{xy}(\mu)$ in units of $10^{4}$~S/m.
    (b) Anomalous Nernst conductivity $\alpha_{xy}(\mu)$ in units of A\,m$^{-1}$\,K$^{-1}$.
    (c) Longitudinal conductivity shown as $\sigma_{yy}(\mu)/\tau$ in units of $10^{20}$~S\,m$^{-1}$\,s$^{-1}$, where $\tau$ is the relaxation time.
    (d) Longitudinal Seebeck coefficient $S_{yy}(\mu)$ in units of $\mu$V/K.
    Panels (c) and (d) provide the longitudinal inputs required to evaluate the transverse thermopower due to ANE, $S_{\mathrm{ANE}}$, via Eq.~\eqref{eq:sane_smallangle}.
    A common color scheme is used across panels to distinguish $\bm{M}\parallel(100)$ and $\bm{M}\parallel(001)$.
    }  \label{fig:DyCo5_calc}
  \end{center}
\end{figure}

\noindent\textbf{Energy-resolved anomalous Hall and Nernst responses.}
Figure \ref{fig:DyCo5_calc} summarizes key transport coefficients of DyCo$_5$ as functions of the chemical potential $\mu$ within a rigid band approximation, where $\mu=0$ corresponds to the Fermi level $E_{\mathrm F}$.
Figure \ref{fig:DyCo5_calc}(a) shows the anomalous Hall conductivity $\sigma_{xy}(\mu)$ for the two magnetization directions.
Both configurations sustain a sizable intrinsic AHC at $\mu=0$, whereas the detailed energy dependence near $E_{\mathrm F}$ differs between $\bm{M}\parallel(001)$ and $\bm{M}\parallel(100)$. 
For reference, the AHC values at $\mu=0$ for the two magnetization directions are summarized in Table~\ref{tab:ahc_M}.
Figure \ref{fig:DyCo5_calc}(b) shows the anomalous Nernst conductivity $\alpha_{xy}(\mu)$ obtained from the same DFT bands and $k$ mesh.
A central observation is that the orientation dependence is modest in $\sigma_{xy}$ but becomes dramatic in $\alpha_{xy}$, reaching about two orders of magnitude contrast near $\mu=0$.
This behavior is consistent with the Mott relation in the low temperature limit, in which $\alpha_{xy}$ is governed by the energy slope $\partial_{\varepsilon}\sigma_{xy}(\varepsilon)\rvert_{E_{\mathrm F}}$ and hence is strongly affected by how Berry curvature is distributed in energy within the thermal window.

In terms of the energy-resolved AHC $\sigma_{xy}(\varepsilon)$, the slope $\partial_{\varepsilon}\sigma_{xy}$ is controlled by the Berry-curvature density at energy $\varepsilon$ (i.e., by how much integrated Berry curvature is accumulated when the chemical potential is moved). Because SOC generates narrow avoided crossings carrying large Berry curvature, even a small magnetization-induced shift of these features by a few meV relative to $E_{\mathrm F}$ can strongly change $\partial_{\varepsilon}\sigma_{xy}\rvert_{E_{\mathrm F}}$ while leaving $\sigma_{xy}(E_{\mathrm F})$ comparatively less affected. This explains why the orientation sensitivity is amplified in $\alpha_{xy}$ compared with $\sigma_{xy}$.

\renewcommand{\arraystretch}{1.5}
\begin{table}
\caption{\label{tab:ahc_M}AHC value at $E_F$ in the (100) and (001) directions.}
\begin{ruledtabular}
\begin{tabular}{ccc}
  & $\mathbf{M}$//(100) & $\mathbf{M}$//(001) \\
\hline
AHC $[\times 10^4 \text{S/m}]$ & 6.18 & 11.80 \\

\end{tabular}
\end{ruledtabular}
\end{table}

\renewcommand{\arraystretch}{1.5}
\begin{table}
\caption{\label{tab:anc_M}$\alpha_{xy}$ value at $E_F$ in the (100) and (001) directions for $T=300$ K and $T=400$ K.}
\begin{ruledtabular}
\begin{tabular}{ccc}
 $\alpha_{xy}$ $[\text{K}^{-1}\text{Am}^{-1}]$ & $\mathbf{M}$//(100) & $\mathbf{M}$//(001) \\
\hline
$T=300$K & 0.0497 & 9.20 \\
$T=400$K & 0.131 & 10.4 \\
\end{tabular}
\end{ruledtabular}
\end{table}

\begin{figure}[htbp]
  \begin{center}
    \includegraphics[width=70mm]{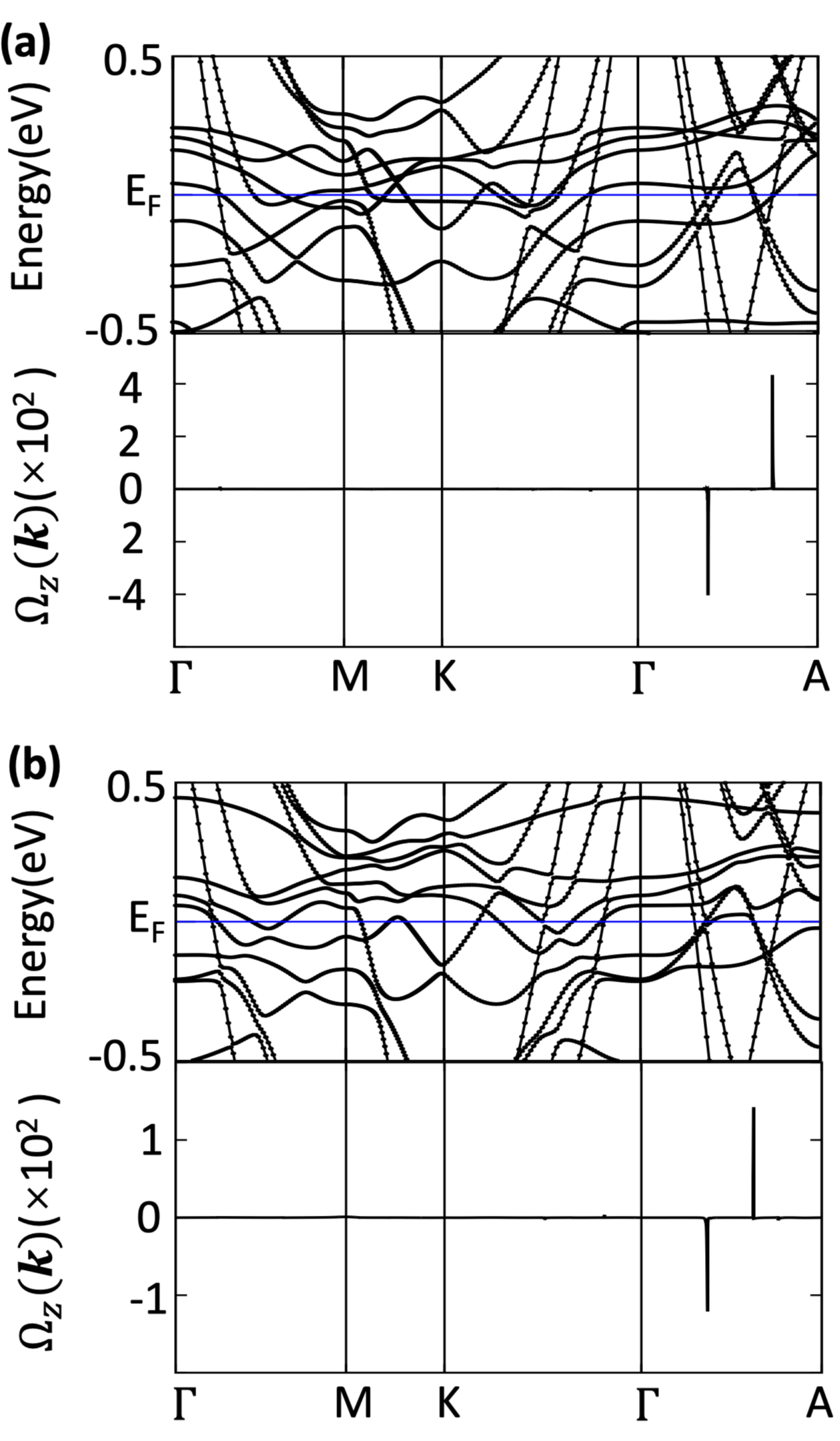}
    \caption{
    Band dispersion and Berry curvature of DyCo$_5$ along the high symmetry path $\Gamma \to M \to K \to \Gamma \to A$ within the energy window $-0.5$~eV $<E<0.5$~eV.
    The upper panels show the band dispersion, while the lower panels plot the Berry curvature summed over occupied states, $\Omega_{z}(\bm{k})=\sum_{n} f(\varepsilon_{n\bm{k}})\,\Omega_{n,z}(\bm{k})$, which corresponds to the integrand of the intrinsic anomalous Hall conductivity up to a constant prefactor.
    (a) $\bm{M}\parallel c$ and (b) $\bm{M}\perp c$.
    In both cases $\Omega_{z}(\bm{k})$ is nearly zero over most of the path but exhibits two sharply peaked hot spots along $\Gamma \to A$, with opposite signs near $\Gamma$ and near $A$.
    Upon reorienting $\bm{M}$, the hot spot positions shift slightly along $\Gamma \to A$, resulting in a modest change in the separation between the two peaks.
    These hot spots coincide with band (avoided) crossings of Co $3d$ states in the dispersion, indicating that the orientation dependence of the thermoelectric Hall response originates from Berry curvature hot spots generated by spin orbit coupling near band degeneracies.
    The hot-spot energies shift upon reorienting $\bm{M}$ because the SOC term $\lambda\,\mathbf{L}\!\cdot\!\mathbf{S}$ couples orbital characters differently for different spin-quantization axes: rotating $\bm{M}$ changes the SOC matrix elements and the pattern of hybridization gaps at near-degenerate Co-$3d$ crossings, thereby moving the avoided-crossing energies (and associated Berry-curvature peaks) relative to $E_{\mathrm F}$.For clarity, this plot is a 1D cut used for visualization; the quantitative $\sigma_{xy}$ and $\alpha_{xy}$ are evaluated from full 3D BZ integrals.} \label{fig:bands_BC}
  \end{center}
\end{figure}

\noindent\textbf{Berry curvature origin of the contrast.}
To identify the microscopic origin of the orientation contrast, Fig.~\ref{fig:bands_BC} presents the band dispersion and the Berry curvature along the high symmetry path $\Gamma \to M \to K \to \Gamma \to A$ for $\bm{M}\parallel c$ and $\bm{M}\perp c$.
While Fig.~\ref{fig:bands_BC} shows a one-dimensional high-symmetry cut for visualization, all reported $\sigma_{xy}$ and $\alpha_{xy}$ values are obtained from full three-dimensional BZ integrals on the dense $k$ mesh.
The Berry curvature shown in the lower panels is the sum over the occupied states,
$\Omega_{z}(\bm{k})=\sum_{n} f(\varepsilon_{n\bm{k}})\,\Omega_{n,z}(\bm{k})$,
which corresponds to the integrand of the intrinsic anomalous Hall conductivity up to a constant prefactor.
In both magnetization configurations, $\Omega_{z}(\bm{k})$ is nearly zero over most of the path but exhibits two strongly peaked hot spots on the $\Gamma \to A$ segment with opposite signs.
These peaks originate from spin orbit coupling gaps at band (avoided) crossings among Co $3d$ derived states, for example involving $d_{z^2}$ with $d_{xz}$ and $d_{x^2-y^2}/d_{xy}$ characters, which generate large $\Omega_{n,z}$.
Upon reorienting $\bm{M}$, the hot spot intensity is redistributed and their positions shift slightly along $\Gamma \to A$, leading to a modest change in the separation between the two peaks.
This redistribution reshapes $\sigma_{xy}(\varepsilon)$ near $E_F$ and hence modifies the energy slope that controls $\alpha_{xy}$, providing a consistent explanation of the trends in Fig.~\ref{fig:DyCo5_calc}(a,b).
A detailed microscopic analysis of the origin of the positional shift is beyond the scope of the present work and is left for future study.

\noindent\textbf{Robustness and consistency checks.}
All qualitative conclusions persist upon (i) modest variations of the onsite interaction on Co-$3d$ (used only as a robustness check), (ii) enlarging the Wannierization window, and (iii) further refining the $k$ mesh beyond $68\times 68\times 73$.
Low $T$ behavior follows the Mott relation, and finite $T$ deviations track nonlinearity in $\sigma_{xy}(\varepsilon)$ within the thermal window.
Device metrics $S_{\mathrm{ANE}}$ and $\Pi_{\mathrm{AEE}}$ are evaluated using longitudinal coefficients from the same band model and a single relaxation time $\tau$ appropriate to a given crystal quality; comparisons are made at a common temperature (300~K).

\noindent\textbf{Room temperature snapshot and implications.}
At $T=300$~K, DyCo$_5$ exhibits a large intrinsic AHC for both magnetization orientations, whereas $\alpha_{xy}$ shows a pronounced orientation contrast under identical conditions.
To provide an order-of-magnitude device estimate, we model a DyCo$_5$ slab of thickness $t$ carrying an in-plane current density $J_c$, for which the AEE generates a transverse heat flux $J_q\approx \Pi_{\mathrm{AEE}} J_c$ (W\,m$^{-2}$).
Using the small-Hall-angle approximation $S_{\mathrm{ANE}}\simeq \rho_{yy}\alpha_{xy}$ (with $\rho_{yy}=1/\sigma_{yy}$) and the Kelvin relation $\Pi_{\mathrm{AEE}}=T S_{\mathrm{ANE}}$, this can also be written as $J_q\approx T\,\alpha_{xy}\,\rho_{yy}\,J_c$.
In steady state, this produces a transverse temperature drop $\Delta T_{\mathrm{AEE}}\sim J_q t/\kappa$, where $\kappa$ is the effective thermal conductivity across the slab. Using the room-temperature values at $\mu=0$ (Table~\ref{tab:anc_M}) together with a representative metallic conductivity $\sigma_{yy}\sim10^{6}$--$10^{7}$~S/m, we obtain $S_{\mathrm{ANE}}\sim \alpha_{xy}/\sigma_{yy}\approx 1$--$10~\mu$V/K in the high-$\alpha_{xy}$ orientation and about two orders of magnitude smaller in the low-$\alpha_{xy}$ orientation, implying $\Pi_{\mathrm{AEE}}\sim0.3$--$3$~mV at 300~K. For current densities $J_c\sim10^{9}$--$10^{10}$~A/m$^{2}$ and $t\sim1$--$10~\mu$m (with $\kappa\sim10$--$30$~W/mK), this corresponds to $\Delta T_{\mathrm{AEE}}\sim0.1$--$10$~K, providing experimentally visible contrast and negative feedback once the device temperature enters the SRT window ($T_{\mathrm{SR1}}\approx325$~K to $T_{\mathrm{SR2}}\approx367$~K).
The resulting transport scales (AHC of order $10^{2}$--$10^{3}$~S/cm and ANE thermopower of order $\mu$V/K) are consistent with reported values for metallic ferromagnets and Co-based magnets, supporting the plausibility of the magnitudes shown in Fig.~\ref{fig:DyCo5_calc}~\cite{Nagaosa2010,Uchida2021}.

Experimentally, the effect can be probed in a DyCo$_5$ thin film patterned into a Hall-bar geometry: an in-plane current $J_c$ is applied while the transverse temperature profile (or heat-flow direction) is detected using micro-thermometry (e.g., resistance thermometers or thermocouples) or scanning thermal/IR thermography on the film surface. Such a film-based implementation directly targets an application scenario of on-chip thermal regulation, where the SRT provides a built-in temperature threshold for autonomous control.

The resulting difference in $S_{\mathrm{ANE}}$ and $\Pi_{\mathrm{AEE}}$ enables a \emph{sensor free but current driven} thermal switch whose state is set by the materials’ internal SRT threshold.
Beyond DyCo$_5$, the mechanism---reorientation driven relocation of Berry curvature hot spots near $E_F$---suggests a practical ``tuning knob'' for $R$Co$_5$ and related ferromagnets: Selecting compositions and strain states that sharpen $\partial_\varepsilon\sigma_{xy}$ near $E_F$ will enhance $\alpha_{xy}$ and improve switching contrast, as long as longitudinal losses remain modest.

%
%
To summarize, we proposed and quantified a {\it{sensor-free, self-regulating}} thermal switch that combines the AEE with a temperature-driven SRT in $R$Co$_5$. Using DFT-Kubo calculations, we established a one-pathway mechanism from the energy-resolved anomalous Hall conductivity $\sigma_{xy}(\varepsilon)$ to the finite-temperature anomalous Nernst conductivity $\alpha_{xy}(T)$. Although the intrinsic AHC at $E_F$ remains sizable for both $\mathbf{M}\parallel c$ and $\mathbf{M}\perp c$, $\alpha_{xy}$ exhibits a pronounced orientation contrast across $T_{\mathrm{SRT}}$. The effect is traced to the \emph{energy slope} $\partial_\varepsilon\sigma_{xy}|_{E_F}$ (consistent with the low-$T$ Mott relation) and ultimately to the relocation/intensity change of \emph{strongly peaked} Berry-curvature hot spots associated with SOC-opened avoided crossings in Co-$3d$ bands. Device-level figures of merit, including the ANE thermopower $S_{\mathrm{ANE}}$ and Ettingshausen coefficient $\Pi_{\mathrm{AEE}}=T\,S_{\mathrm{ANE}}$, show robust switching at a fixed temperature (e.g., 300~K) under a constant in-plane bias current, enabling compact, current-driven, sensor-free thermal control.

The materials-level thresholding demonstrated in DyCo$_5$ provides a practical design parameter for $R$Co$_5$ and related ferromagnets: Compositions, strain states, and anisotropy engineering that sharpen $\partial_\varepsilon\sigma_{xy}$ near $E_F$ are expected to enhance $\alpha_{xy}$ and improve switching contrast, provided longitudinal losses remain modest. 
%
%

\bibliography{APL}

\begin{thebibliography}{26}%
\makeatletter
\providecommand \@ifxundefined [1]{%
 \@ifx{#1\undefined}
}%
\providecommand \@ifnum [1]{%
 \ifnum #1\expandafter \@firstoftwo
 \else \expandafter \@secondoftwo
 \fi
}%
\providecommand \@ifx [1]{%
 \ifx #1\expandafter \@firstoftwo
 \else \expandafter \@secondoftwo
 \fi
}%
\providecommand \natexlab [1]{#1}%
\providecommand \enquote  [1]{``#1''}%
\providecommand \bibnamefont  [1]{#1}%
\providecommand \bibfnamefont [1]{#1}%
\providecommand \citenamefont [1]{#1}%
\providecommand \href@noop [0]{\@secondoftwo}%
\providecommand \href [0]{\begingroup \@sanitize@url \@href}%
\providecommand \@href[1]{\@@startlink{#1}\@@href}%
\providecommand \@@href[1]{\endgroup#1\@@endlink}%
\providecommand \@sanitize@url [0]{\catcode `\\12\catcode `\$12\catcode `\&12\catcode `\#12\catcode `\^12\catcode `\_12\catcode `\%12\relax}%
\providecommand \@@startlink[1]{}%
\providecommand \@@endlink[0]{}%
\providecommand \url  [0]{\begingroup\@sanitize@url \@url }%
\providecommand \@url [1]{\endgroup\@href {#1}{\urlprefix }}%
\providecommand \urlprefix  [0]{URL }%
\providecommand \Eprint [0]{\href }%
\providecommand \doibase [0]{http://dx.doi.org/}%
\providecommand \selectlanguage [0]{\@gobble}%
\providecommand \bibinfo  [0]{\@secondoftwo}%
\providecommand \bibfield  [0]{\@secondoftwo}%
\providecommand \translation [1]{[#1]}%
\providecommand \BibitemOpen [0]{}%
\providecommand \bibitemStop [0]{}%
\providecommand \bibitemNoStop [0]{.\EOS\space}%
\providecommand \EOS [0]{\spacefactor3000\relax}%
\providecommand \BibitemShut  [1]{\csname bibitem#1\endcsname}%
\let\auto@bib@innerbib\@empty
\bibitem [{\citenamefont {Pop}(2010)}]{Pop2010}%
  \BibitemOpen
  \bibfield  {author} {\bibinfo {author} {\bibfnamefont {E.}~\bibnamefont {Pop}},\ }\href {\doibase 10.1007/s12274-010-1019-z} {\bibfield  {journal} {\bibinfo  {journal} {Nano Res.}\ }\textbf {\bibinfo {volume} {3}},\ \bibinfo {pages} {147--169} (\bibinfo {year} {2010})}\BibitemShut {NoStop}%
\bibitem [{\citenamefont {Moore}\ and\ \citenamefont {Shi}(2010)}]{Moore2010}%
  \BibitemOpen
  \bibfield  {author} {\bibinfo {author} {\bibfnamefont {A.~L.}\ \bibnamefont {Moore}}\ and\ \bibinfo {author} {\bibfnamefont {L.}~\bibnamefont {Shi}},\ }\href {\doibase 10.1016/j.mattod.2014.04.003} {\bibfield  {journal} {\bibinfo  {journal} {Materials Today}\ }\textbf {\bibinfo {volume} {17}},\ \bibinfo {pages} {163--174} (\bibinfo {year} {2010})}\BibitemShut {NoStop}%
\bibitem [{\citenamefont {Uchida}, \citenamefont {Zhou},\ and\ \citenamefont {Sakuraba}(2021)}]{Uchida2021}%
  \BibitemOpen
  \bibfield  {author} {\bibinfo {author} {\bibfnamefont {K.-i.}\ \bibnamefont {Uchida}}, \bibinfo {author} {\bibfnamefont {W.}~\bibnamefont {Zhou}}, \ and\ \bibinfo {author} {\bibfnamefont {Y.}~\bibnamefont {Sakuraba}},\ }\href {\doibase 10.1063/5.0046877} {\bibfield  {journal} {\bibinfo  {journal} {Appl. Phys. Lett.}\ }\textbf {\bibinfo {volume} {18}},\ \bibinfo {pages} {140504} (\bibinfo {year} {2021})}\BibitemShut {NoStop}%
\bibitem [{\citenamefont {Nernst}(1887)}]{Nernst1887}%
  \BibitemOpen
  \bibfield  {author} {\bibinfo {author} {\bibfnamefont {W.}~\bibnamefont {Nernst}},\ }\href {\doibase 10.1002/andp.18872670815} {\bibfield  {journal} {\bibinfo  {journal} {Annalen der Physik}\ }\textbf {\bibinfo {volume} {267}},\ \bibinfo {pages} {760--789} (\bibinfo {year} {1887})}\BibitemShut {NoStop}%
\bibitem [{\citenamefont {V.~Ettingshausen}\ and\ \citenamefont {Nernst}(1886)}]{Etting1886}%
  \BibitemOpen
  \bibfield  {author} {\bibinfo {author} {\bibfnamefont {A.}~\bibnamefont {V.~Ettingshausen}}\ and\ \bibinfo {author} {\bibfnamefont {W.}~\bibnamefont {Nernst}},\ }\href {\doibase 10.1002/andp.18862651010} {\bibfield  {journal} {\bibinfo  {journal} {Annalen der Physik}\ }\textbf {\bibinfo {volume} {265}},\ \bibinfo {pages} {343--347} (\bibinfo {year} {1886})}\BibitemShut {NoStop}%
\bibitem [{\citenamefont {Nagaosa}\ \emph {et~al.}(2010)\citenamefont {Nagaosa}, \citenamefont {Sinova}, \citenamefont {Onoda}, \citenamefont {MacDonald},\ and\ \citenamefont {Ong}}]{Nagaosa2010}%
  \BibitemOpen
  \bibfield  {author} {\bibinfo {author} {\bibfnamefont {N.}~\bibnamefont {Nagaosa}}, \bibinfo {author} {\bibfnamefont {J.}~\bibnamefont {Sinova}}, \bibinfo {author} {\bibfnamefont {S.}~\bibnamefont {Onoda}}, \bibinfo {author} {\bibfnamefont {A.~H.}\ \bibnamefont {MacDonald}}, \ and\ \bibinfo {author} {\bibfnamefont {N.~P.}\ \bibnamefont {Ong}},\ }\href {\doibase 10.1103/RevModPhys.82.1539} {\bibfield  {journal} {\bibinfo  {journal} {Rev. Mod. Phys.}\ }\textbf {\bibinfo {volume} {82}},\ \bibinfo {pages} {1539--1592} (\bibinfo {year} {2010})}\BibitemShut {NoStop}%
\bibitem [{\citenamefont {Miura}\ \emph {et~al.}(2019)\citenamefont {Miura}, \citenamefont {Sepehri-Amin}, \citenamefont {Masuda}, \citenamefont {Tsuchiura}, \citenamefont {Miura}, \citenamefont {Iguchi}, \citenamefont {Sakuraba}, \citenamefont {Shiomi}, \citenamefont {Hono},\ and\ \citenamefont {Uchida}}]{Miura2019}%
  \BibitemOpen
  \bibfield  {author} {\bibinfo {author} {\bibfnamefont {A.}~\bibnamefont {Miura}}, \bibinfo {author} {\bibfnamefont {H.}~\bibnamefont {Sepehri-Amin}}, \bibinfo {author} {\bibfnamefont {K.}~\bibnamefont {Masuda}}, \bibinfo {author} {\bibfnamefont {H.}~\bibnamefont {Tsuchiura}}, \bibinfo {author} {\bibfnamefont {Y.}~\bibnamefont {Miura}}, \bibinfo {author} {\bibfnamefont {R.}~\bibnamefont {Iguchi}}, \bibinfo {author} {\bibfnamefont {Y.}~\bibnamefont {Sakuraba}}, \bibinfo {author} {\bibfnamefont {J.}~\bibnamefont {Shiomi}}, \bibinfo {author} {\bibfnamefont {K.}~\bibnamefont {Hono}}, \ and\ \bibinfo {author} {\bibfnamefont {K.}~\bibnamefont {Uchida}},\ }\href {\doibase 10.1063/1.5131001} {\bibfield  {journal} {\bibinfo  {journal} {Appl. Phys. Lett.}\ }\textbf {\bibinfo {volume} {115}},\ \bibinfo {pages} {222403} (\bibinfo {year} {2019})}\BibitemShut {NoStop}%
\bibitem [{\citenamefont {Tsushima}\ and\ \citenamefont {Ohokoshi}(1983)}]{Tsushima1983}%
  \BibitemOpen
  \bibfield  {author} {\bibinfo {author} {\bibfnamefont {T.}~\bibnamefont {Tsushima}}\ and\ \bibinfo {author} {\bibfnamefont {M.}~\bibnamefont {Ohokoshi}},\ }\href {\doibase 10.1016/0304-8853(83)90213-5} {\bibfield  {journal} {\bibinfo  {journal} {Journal of Magnetism and Magnetic Materials}\ }\textbf {\bibinfo {volume} {31-34}},\ \bibinfo {pages} {197} (\bibinfo {year} {1983})}\BibitemShut {NoStop}%
\bibitem [{\citenamefont {Vishina}\ \emph {et~al.}(2025)\citenamefont {Vishina}, \citenamefont {Skokov}, \citenamefont {Tsuchiura}, \citenamefont {Thunstr\"{o}m}, \citenamefont {Aubert}, \citenamefont {Gutfleisch}, \citenamefont {Eriksson},\ and\ \citenamefont {Herper}}]{Alena2025}%
  \BibitemOpen
  \bibfield  {author} {\bibinfo {author} {\bibfnamefont {A.}~\bibnamefont {Vishina}}, \bibinfo {author} {\bibfnamefont {K.}~\bibnamefont {Skokov}}, \bibinfo {author} {\bibfnamefont {H.}~\bibnamefont {Tsuchiura}}, \bibinfo {author} {\bibfnamefont {P.}~\bibnamefont {Thunstr\"{o}m}}, \bibinfo {author} {\bibfnamefont {A.}~\bibnamefont {Aubert}}, \bibinfo {author} {\bibfnamefont {O.}~\bibnamefont {Gutfleisch}}, \bibinfo {author} {\bibfnamefont {O.}~\bibnamefont {Eriksson}}, \ and\ \bibinfo {author} {\bibfnamefont {H.~C.}\ \bibnamefont {Herper}},\ }\href@noop {} {\bibfield  {journal} {\bibinfo  {journal} {arXiv:2511.17087}\ } (\bibinfo {year} {2025})}\BibitemShut {NoStop}%
\bibitem [{\citenamefont {Ohkoshi}\ \emph {et~al.}(1977)\citenamefont {Ohkoshi}, \citenamefont {Kobayashi}, \citenamefont {Katayama}, \citenamefont {Hirano},\ and\ \citenamefont {Tsushima}}]{Ohkoshi1977}%
  \BibitemOpen
  \bibfield  {author} {\bibinfo {author} {\bibfnamefont {M.}~\bibnamefont {Ohkoshi}}, \bibinfo {author} {\bibfnamefont {H.}~\bibnamefont {Kobayashi}}, \bibinfo {author} {\bibfnamefont {T.}~\bibnamefont {Katayama}}, \bibinfo {author} {\bibfnamefont {M.}~\bibnamefont {Hirano}}, \ and\ \bibinfo {author} {\bibfnamefont {T.}~\bibnamefont {Tsushima}},\ }\href {\doibase 10.1016/0378-4363(77)90284-4} {\bibfield  {journal} {\bibinfo  {journal} {Physica B+C}\ }\textbf {\bibinfo {volume} {86}},\ \bibinfo {pages} {195} (\bibinfo {year} {1977})}\BibitemShut {NoStop}%
\bibitem [{\citenamefont {Tereshina}\ \emph {et~al.}(2007)\citenamefont {Tereshina}, \citenamefont {Korenovskii}, \citenamefont {Burkhanov}, \citenamefont {Kuz’min}, \citenamefont {Skokov},\ and\ \citenamefont {Melero}}]{Tereshina2007}%
  \BibitemOpen
  \bibfield  {author} {\bibinfo {author} {\bibfnamefont {I.~S.}\ \bibnamefont {Tereshina}}, \bibinfo {author} {\bibfnamefont {N.~L.}\ \bibnamefont {Korenovskii}}, \bibinfo {author} {\bibfnamefont {G.~S.}\ \bibnamefont {Burkhanov}}, \bibinfo {author} {\bibfnamefont {M.~D.}\ \bibnamefont {Kuz’min}}, \bibinfo {author} {\bibfnamefont {K.~P.}\ \bibnamefont {Skokov}}, \ and\ \bibinfo {author} {\bibfnamefont {J.~J.}\ \bibnamefont {Melero}},\ }\href {\doibase 10.1134/S106377610712014X} {\bibfield  {journal} {\bibinfo  {journal} {Journal of Experimental and Theoretical Physics}\ }\textbf {\bibinfo {volume} {105}},\ \bibinfo {pages} {1230} (\bibinfo {year} {2007})}\BibitemShut {NoStop}%
\bibitem [{\citenamefont {Belov}\ \emph {et~al.}(1976)\citenamefont {Belov}, \citenamefont {Zvezdin}, \citenamefont {Kadomtsewa},\ and\ \citenamefont {Levitin}}]{Belov1976}%
  \BibitemOpen
  \bibfield  {author} {\bibinfo {author} {\bibfnamefont {K.~P.}\ \bibnamefont {Belov}}, \bibinfo {author} {\bibfnamefont {A.~K.}\ \bibnamefont {Zvezdin}}, \bibinfo {author} {\bibfnamefont {A.~M.}\ \bibnamefont {Kadomtsewa}}, \ and\ \bibinfo {author} {\bibfnamefont {R.~Z.}\ \bibnamefont {Levitin}},\ }\href {\doibase 10.1070/PU1976v019n07ABEH005274} {\bibfield  {journal} {\bibinfo  {journal} {Sov. Phys. Usp.}\ }\textbf {\bibinfo {volume} {19}},\ \bibinfo {pages} {574} (\bibinfo {year} {1976})}\BibitemShut {NoStop}%
\bibitem [{\citenamefont {Dorantes-D\'avila}, \citenamefont {Garibay-Alonso},\ and\ \citenamefont {Pastor}(2024)}]{Davila2024}%
  \BibitemOpen
  \bibfield  {author} {\bibinfo {author} {\bibfnamefont {J.}~\bibnamefont {Dorantes-D\'avila}}, \bibinfo {author} {\bibfnamefont {R.}~\bibnamefont {Garibay-Alonso}}, \ and\ \bibinfo {author} {\bibfnamefont {G.~M.}\ \bibnamefont {Pastor}},\ }\href {\doibase 10.1103/PhysRevB.110.174406} {\bibfield  {journal} {\bibinfo  {journal} {Phys. Rev. B}\ }\textbf {\bibinfo {volume} {110}},\ \bibinfo {pages} {174406} (\bibinfo {year} {2024})}\BibitemShut {NoStop}%
\bibitem [{\citenamefont {Mott}\ and\ \citenamefont {Jones}(1936)}]{Mott1936}%
  \BibitemOpen
  \bibfield  {author} {\bibinfo {author} {\bibfnamefont {N.~F.}\ \bibnamefont {Mott}}\ and\ \bibinfo {author} {\bibfnamefont {H.}~\bibnamefont {Jones}},\ }\href@noop {} {\bibfield  {journal} {\bibinfo  {journal} {Theory of the Properties of Metals and Alloys. Oxford University Press, Reprinted by Dover Publications, New York.}\ } (\bibinfo {year} {1936})}\BibitemShut {NoStop}%
\bibitem [{\citenamefont {Cutler}\ and\ \citenamefont {Mott}(1969)}]{Mott1969}%
  \BibitemOpen
  \bibfield  {author} {\bibinfo {author} {\bibfnamefont {M.}~\bibnamefont {Cutler}}\ and\ \bibinfo {author} {\bibfnamefont {N.~F.}\ \bibnamefont {Mott}},\ }\href {\doibase https://doi.org/10.1103/PhysRev.181.1336} {\bibfield  {journal} {\bibinfo  {journal} {Phys. Rev.}\ }\textbf {\bibinfo {volume} {181}},\ \bibinfo {pages} {1336} (\bibinfo {year} {1969})}\BibitemShut {NoStop}%
\bibitem [{\citenamefont {Seki}\ \emph {et~al.}(2018)\citenamefont {Seki}, \citenamefont {Iguchi}, \citenamefont {Takanashi},\ and\ \citenamefont {Uchida}}]{Seki2018}%
  \BibitemOpen
  \bibfield  {author} {\bibinfo {author} {\bibfnamefont {T.}~\bibnamefont {Seki}}, \bibinfo {author} {\bibfnamefont {R.}~\bibnamefont {Iguchi}}, \bibinfo {author} {\bibfnamefont {K.}~\bibnamefont {Takanashi}}, \ and\ \bibinfo {author} {\bibfnamefont {K.}~\bibnamefont {Uchida}},\ }\href {\doibase 10.1088/1361-6463/aac481} {\bibfield  {journal} {\bibinfo  {journal} {J. Phys. D: Appl. Phys.}\ }\textbf {\bibinfo {volume} {51}},\ \bibinfo {pages} {254001} (\bibinfo {year} {2018})}\BibitemShut {NoStop}%
\bibitem [{\citenamefont {Blaha}\ \emph {et~al.}(2001)\citenamefont {Blaha}, \citenamefont {Schwarz}, \citenamefont {Madsen}, \citenamefont {Kvasnicka}, \citenamefont {Luitz} \emph {et~al.}}]{blaha2001wien2k}%
  \BibitemOpen
  \bibfield  {author} {\bibinfo {author} {\bibfnamefont {P.}~\bibnamefont {Blaha}}, \bibinfo {author} {\bibfnamefont {K.}~\bibnamefont {Schwarz}}, \bibinfo {author} {\bibfnamefont {G.~K.}\ \bibnamefont {Madsen}}, \bibinfo {author} {\bibfnamefont {D.}~\bibnamefont {Kvasnicka}}, \bibinfo {author} {\bibfnamefont {J.}~\bibnamefont {Luitz}},  \emph {et~al.},\ }\href@noop {} {\bibfield  {journal} {\bibinfo  {journal} {An augmented plane wave+ local orbitals program for calculating crystal properties}\ }\textbf {\bibinfo {volume} {60}} (\bibinfo {year} {2001})}\BibitemShut {NoStop}%
\bibitem [{\citenamefont {Perdew}, \citenamefont {Burke},\ and\ \citenamefont {Ernzerhof}(1996)}]{Perdew1996}%
  \BibitemOpen
  \bibfield  {author} {\bibinfo {author} {\bibfnamefont {J.~P.}\ \bibnamefont {Perdew}}, \bibinfo {author} {\bibfnamefont {K.}~\bibnamefont {Burke}}, \ and\ \bibinfo {author} {\bibfnamefont {M.}~\bibnamefont {Ernzerhof}},\ }\href {\doibase 10.1103/PhysRevLett.77.3865} {\bibfield  {journal} {\bibinfo  {journal} {Phys. Rev. Lett.}\ }\textbf {\bibinfo {volume} {77}},\ \bibinfo {pages} {3865} (\bibinfo {year} {1996})}\BibitemShut {NoStop}%
\bibitem [{\citenamefont {liechtenstein}, \citenamefont {Anisimov},\ and\ \citenamefont {Zaanen}(1995)}]{Liechtenstein1995}%
  \BibitemOpen
  \bibfield  {author} {\bibinfo {author} {\bibfnamefont {A.~I.}\ \bibnamefont {liechtenstein}}, \bibinfo {author} {\bibfnamefont {V.~I.}\ \bibnamefont {Anisimov}}, \ and\ \bibinfo {author} {\bibfnamefont {J.}~\bibnamefont {Zaanen}},\ }\href {\doibase 10.1103/PhysRevB.52.R5467} {\bibfield  {journal} {\bibinfo  {journal} {Phys. Rev. B}\ }\textbf {\bibinfo {volume} {52}},\ \bibinfo {pages} {R5467--R5470} (\bibinfo {year} {1995})}\BibitemShut {NoStop}%
\bibitem [{\citenamefont {Pizzi}\ \emph {et~al.}(2020)\citenamefont {Pizzi}, \citenamefont {Vitale}, \citenamefont {Arita}, \citenamefont {Bl{\"u}gel}, \citenamefont {Freimuth}, \citenamefont {G{\'e}ranton}, \citenamefont {Gibertini}, \citenamefont {Gresch}, \citenamefont {Johnson}, \citenamefont {Koretsune} \emph {et~al.}}]{Pizzi2020}%
  \BibitemOpen
  \bibfield  {author} {\bibinfo {author} {\bibfnamefont {G.}~\bibnamefont {Pizzi}}, \bibinfo {author} {\bibfnamefont {V.}~\bibnamefont {Vitale}}, \bibinfo {author} {\bibfnamefont {R.}~\bibnamefont {Arita}}, \bibinfo {author} {\bibfnamefont {S.}~\bibnamefont {Bl{\"u}gel}}, \bibinfo {author} {\bibfnamefont {F.}~\bibnamefont {Freimuth}}, \bibinfo {author} {\bibfnamefont {G.}~\bibnamefont {G{\'e}ranton}}, \bibinfo {author} {\bibfnamefont {M.}~\bibnamefont {Gibertini}}, \bibinfo {author} {\bibfnamefont {D.}~\bibnamefont {Gresch}}, \bibinfo {author} {\bibfnamefont {C.}~\bibnamefont {Johnson}}, \bibinfo {author} {\bibfnamefont {T.}~\bibnamefont {Koretsune}},  \emph {et~al.},\ }\href {\doibase 10.1088/1361-648X/ab51ff} {\bibfield  {journal} {\bibinfo  {journal} {Journal of Physics: Condensed Matter}\ }\textbf {\bibinfo {volume} {32}},\ \bibinfo {pages} {165902} (\bibinfo {year} {2020})}\BibitemShut {NoStop}%
\bibitem [{\citenamefont {Marzari}\ \emph {et~al.}(2012)\citenamefont {Marzari}, \citenamefont {Mostofi}, \citenamefont {Yates}, \citenamefont {Souza},\ and\ \citenamefont {Vanderbilt}}]{Marzari2012}%
  \BibitemOpen
  \bibfield  {author} {\bibinfo {author} {\bibfnamefont {N.}~\bibnamefont {Marzari}}, \bibinfo {author} {\bibfnamefont {A.~A.}\ \bibnamefont {Mostofi}}, \bibinfo {author} {\bibfnamefont {J.~R.}\ \bibnamefont {Yates}}, \bibinfo {author} {\bibfnamefont {I.}~\bibnamefont {Souza}}, \ and\ \bibinfo {author} {\bibfnamefont {D.}~\bibnamefont {Vanderbilt}},\ }\href {\doibase 10.1103/RevModPhys.84.1419} {\bibfield  {journal} {\bibinfo  {journal} {Rev. Mod. Phys.}\ }\textbf {\bibinfo {volume} {84}},\ \bibinfo {pages} {1419} (\bibinfo {year} {2012})}\BibitemShut {NoStop}%
\bibitem [{\citenamefont {Madsen}, \citenamefont {Carrete},\ and\ \citenamefont {Verstraete}(2018)}]{Madsen2018}%
  \BibitemOpen
  \bibfield  {author} {\bibinfo {author} {\bibfnamefont {G.~K.~H.}\ \bibnamefont {Madsen}}, \bibinfo {author} {\bibfnamefont {J.}~\bibnamefont {Carrete}}, \ and\ \bibinfo {author} {\bibfnamefont {M.~J.}\ \bibnamefont {Verstraete}},\ }\href {\doibase 10.1016/j.cpc.2018.05.010} {\bibfield  {journal} {\bibinfo  {journal} {Comput. Phys. Commun.}\ }\textbf {\bibinfo {volume} {231}},\ \bibinfo {pages} {140 -- 145} (\bibinfo {year} {2018})}\BibitemShut {NoStop}%
\bibitem [{\citenamefont {Xiao}\ \emph {et~al.}(2006)\citenamefont {Xiao}, \citenamefont {Yao}, \citenamefont {Fang},\ and\ \citenamefont {Niu}}]{Xiao2006}%
  \BibitemOpen
  \bibfield  {author} {\bibinfo {author} {\bibfnamefont {D.}~\bibnamefont {Xiao}}, \bibinfo {author} {\bibfnamefont {Y.}~\bibnamefont {Yao}}, \bibinfo {author} {\bibfnamefont {Z.}~\bibnamefont {Fang}}, \ and\ \bibinfo {author} {\bibfnamefont {Q.}~\bibnamefont {Niu}},\ }\href {\doibase 10.1103/PhysRevLett.97.026603} {\bibfield  {journal} {\bibinfo  {journal} {Phys. Rev. Lett.}\ }\textbf {\bibinfo {volume} {97}},\ \bibinfo {pages} {026603} (\bibinfo {year} {2006})}\BibitemShut {NoStop}%
\bibitem [{\citenamefont {Xiao}, \citenamefont {Chang},\ and\ \citenamefont {Niu}(2010)}]{Xiao2010}%
  \BibitemOpen
  \bibfield  {author} {\bibinfo {author} {\bibfnamefont {D.}~\bibnamefont {Xiao}}, \bibinfo {author} {\bibfnamefont {M.-C.}\ \bibnamefont {Chang}}, \ and\ \bibinfo {author} {\bibfnamefont {Q.}~\bibnamefont {Niu}},\ }\href {\doibase 10.1103/RevModPhys.82.1959} {\bibfield  {journal} {\bibinfo  {journal} {Rev. Mod. Phys.}\ }\textbf {\bibinfo {volume} {82}},\ \bibinfo {pages} {1959--2007} (\bibinfo {year} {2010})}\BibitemShut {NoStop}%
\bibitem [{\citenamefont {Ikhlas}\ \emph {et~al.}(2017)\citenamefont {Ikhlas}, \citenamefont {Tomita}, \citenamefont {Koretsune}, \citenamefont {Suzuki}, \citenamefont {Nishio-Hamane}, \citenamefont {Arita}, \citenamefont {Otani},\ and\ \citenamefont {Nakatsuji}}]{Ikhlas2017}%
  \BibitemOpen
  \bibfield  {author} {\bibinfo {author} {\bibfnamefont {M.}~\bibnamefont {Ikhlas}}, \bibinfo {author} {\bibfnamefont {T.}~\bibnamefont {Tomita}}, \bibinfo {author} {\bibfnamefont {T.}~\bibnamefont {Koretsune}}, \bibinfo {author} {\bibfnamefont {M.-T.}\ \bibnamefont {Suzuki}}, \bibinfo {author} {\bibfnamefont {D.}~\bibnamefont {Nishio-Hamane}}, \bibinfo {author} {\bibfnamefont {R.}~\bibnamefont {Arita}}, \bibinfo {author} {\bibfnamefont {Y.}~\bibnamefont {Otani}}, \ and\ \bibinfo {author} {\bibfnamefont {S.}~\bibnamefont {Nakatsuji}},\ }\href {\doibase 10.1038/nphys4181} {\bibfield  {journal} {\bibinfo  {journal} {Nature Physics}\ }\textbf {\bibinfo {volume} {13}},\ \bibinfo {pages} {1085--1090} (\bibinfo {year} {2017})}\BibitemShut {NoStop}%
\bibitem [{\citenamefont {Seifert}\ \emph {et~al.}(2021)\citenamefont {Seifert}, \citenamefont {Martens}, \citenamefont {Radu}, \citenamefont {Ribow}, \citenamefont {Berritta}, \citenamefont {Nádvorník}, \citenamefont {Starke}, \citenamefont {Jungwirth}, \citenamefont {Wolf}, \citenamefont {Radu}, \citenamefont {Münzenberg}, \citenamefont {Oppeneer}, \citenamefont {Woltersdorf},\ and\ \citenamefont {Kampfrath}}]{Seifert2021}%
  \BibitemOpen
  \bibfield  {author} {\bibinfo {author} {\bibfnamefont {T.~S.}\ \bibnamefont {Seifert}}, \bibinfo {author} {\bibfnamefont {U.}~\bibnamefont {Martens}}, \bibinfo {author} {\bibfnamefont {F.}~\bibnamefont {Radu}}, \bibinfo {author} {\bibfnamefont {M.}~\bibnamefont {Ribow}}, \bibinfo {author} {\bibfnamefont {M.}~\bibnamefont {Berritta}}, \bibinfo {author} {\bibfnamefont {L.}~\bibnamefont {Nádvorník}}, \bibinfo {author} {\bibfnamefont {R.}~\bibnamefont {Starke}}, \bibinfo {author} {\bibfnamefont {T.}~\bibnamefont {Jungwirth}}, \bibinfo {author} {\bibfnamefont {M.}~\bibnamefont {Wolf}}, \bibinfo {author} {\bibfnamefont {I.}~\bibnamefont {Radu}}, \bibinfo {author} {\bibfnamefont {M.}~\bibnamefont {Münzenberg}}, \bibinfo {author} {\bibfnamefont {P.~M.}\ \bibnamefont {Oppeneer}}, \bibinfo {author} {\bibfnamefont {G.}~\bibnamefont {Woltersdorf}}, \ and\ \bibinfo {author} {\bibfnamefont {T.}~\bibnamefont {Kampfrath}},\ }\href {\doibase 10.1002/adma.202007398} {\bibfield  {journal} {\bibinfo  {journal} {Adv.
  Mater.}\ }\textbf {\bibinfo {volume} {33}},\ \bibinfo {pages} {2007398} (\bibinfo {year} {2021})}\BibitemShut {NoStop}%
\end{thebibliography}%

\end{document}